\documentclass[traditabstract,rnote]{aa}

\usepackage{graphicx}
\usepackage{txfonts}
\usepackage{epsfig}
\usepackage{natbib}
\bibpunct{(}{)}{;}{a}{}{,}

\begin{document}

\title{Redshift constraints for RGB 0136+391 and PKS 0735+178 from deep optical
imaging\thanks{Based on observations made with the Nordic Optical Telescope,
operated on the island of La Palma jointly by Denmark, Finland, Iceland,
Norway, and Sweden, in the Spanish Observatorio del Roque de los
Muchachos of the Instituto de Astrofisica de Canarias.}}

\author{
K. Nilsson\inst{1} \and
T. Pursimo\inst{2} \and
C. Villforth\inst{3} \and
E. Lindfors\inst{4} \and
L. O. Takalo\inst{4} \and
A. Sillanp\"a\"a\inst{4}
}

\institute{
Finnish Centre for Astronomy with ESO (FINCA), University of Turku,
V\"ais\"al\"antie 20, FI-21500 Piikki\"o, Finland
\and
Nordic Optical Telescope, Apartado 474, 38700 Santa Cruz de La
Palma, Spain
\and
Department of Astronomy, University of Florida, 211 Bryant Space Science
Center, Gainesville, FL 32611-2055, USA
\and
Tuorla Observatory, Department of Physics and Astronomy, University
of Turku, V\"ais\"al\"antie 20, FI-21500 Piikki\"o, Finland
}

\date{Received; accepted}

\abstract{ We present the results of deep I-band imaging of two BL
  Lacerate objects, RGB 0136+391 and PKS 0735+178, during an epoch
  when the optical nucleus was in a faint state in both targets.  In
  PKS~0735+178 we find a significant excess over a point source,
  which, if fitted by the de Vaucouleurs model, corresponds to a
  galaxy with $\rm I = 18.64 \pm 0.11$ and $\rm r_{eff} = 1.8 \pm 0.4$
  arcsec. Interpreting this galaxy as the host galaxy of PKS~0735+178
  we derive z = $0.45 \pm 0.06$ using the host galaxy as a ``standard
  candle''. We also discuss the immediate optical environment of
  PKS~0735+178 and the identity of the MgII absorber at z =
  0.424. Despite of the optimally chosen epoch and deep imaging we
  find the surface brightness profile of RGB 0136+391 to be consistent
  with a point source. By determining a lower limit for the host
  galaxy brightness by simulations, we derive $z > 0.40$ for this
  target.}


\maketitle

\section{Introduction}

The optical spectrum of BL Lacertae objects (BL Lacs) is dominated by
a featureless synchrotron continuum, which makes the determination of
their redshift by spectroscopy very challenging. Redshift
determination of quasars and BL Lacs is usually based on emission
lines originating in the narrow and broad line regions and/or on the
absorption features of the host galaxy. However, in BL Lacs the
emission lines are intrinsically weak and like the host galaxy
absorption lines they often become so weak relative to the continuum
that they fall below the noise level and become unrecognizable.

It is therefore not surprising that many BL Lac samples still have
a high fraction of targets without spectroscopically determined
redshifts. For instance, in one of the most studied samples, the 1 Jy
sample \citep{1991ApJ...374..431S}, 23 of 37 targets have a
spectroscopically secured redshift with the remaining being unknown,
lower limits from an absorption line system or uncertain due to a
detection of a single line \citep{2001AJ....122..565R}. Large (8 m and
up) telescopes are often needed to secure the redshift
\citep[e.g.][]{2009AJ....137..337S}, but even then there often
remains a fraction of targets with unknown redshifts.

\cite{2005ApJ...635..173S} discussed in detail the method of
estimating the redshift of BL Lacs through detection of their host
galaxies. They showed that the absolute magnitude distribution of BL
Lac host galaxies is roughly Gaussian with an average of $M_{\rm
  R}^{\rm host}$ = -22.8 and $\sigma$ = 0.5 mag. This opens the
possibility of estimating the redshift using the host galaxy as a
``standard candle'' if sufficiently deep optical or NIR images can be
obtained. This method has been applied to several BL Lacs to estimate
their redshifts \citep{2008A&A...487L..29N,2010ApJ...712...14M}. Given
the width of the host galaxy luminosity distribution, the imaging
redshifts are not as accurate as spectroscopic redshifts. The accuracy
of the method is $\pm$0.05 in redshift, if high signal-to-noise data
are available, but in practice the redshift error is usually larger
due to noise in derived host galaxy properties
\citep{2005ApJ...635..173S}.

In this paper we report the results of deep optical imaging of two
$\gamma$-ray bright BL Lacs, RGB~0136+391 and PKS~0735+178, which were
detected by the {\it Fermi}-LAT already during first three months of
operations \citep{2009ApJ...700..597A}. For $\gamma$-ray emitting BL
Lac objects the redshift estimation is particularly important and
there are large dedicated spectroscopy programs for determining
redshift for {\it Fermi}-LAT detected AGN. Still 56\% of the LAT
detected BL Lac objects have no redshift measurement
\citep{2011ApJ...743..171A}, which makes population studies difficult
. RGB~0136+391 is particularly interesting, because of its very hard
$\gamma$-ray spectrum (spectral index 1.69) in the Fermi-LAT range and
the very recent discovery of VHE $\gamma$-rays from the source
\citep{MazinProceedings}.  VHE $\gamma$-rays do not travel unaffected
in the intergalactic space, but are absorbed by the extragalactic
background light (EBL). The EBL leaves an imprint on the observed
energy spectrum and if the redshift of the emitting object is known,
the observed energy spectrum can be used to set constrains to density
and evolution of EBL with redshift \citep[e.g.][]{1996ApJ...473L..75S,
  2006Natur.440.1018A,
  2007A&A...471..439M,2012arXiv1202.2867M}. Alternatively, if the
redshift is not known the observed VHE $\gamma$-ray spectrum combined
with the GeV range spectra can also be used to derive independent
estimation for the redshift of the object \citep{2010MNRAS.405L..76P,
  2011arXiv1111.0913P}.

RGB~0136+391 was first identified as a BL Lac object by
\citet{1998ApJS..118..127L} who saw a featureless optical spectrum.
Following attempts 
\citep{1999A&AS..139..575W,2000ApJS..129..547B,2002MNRAS.329..877C,
2007A&A...470..787P} have all failed to detect any spectral features
and the redshift remains unknown. \citet{2007A&A...470..787P} derived a
lower limit z $>$ 0.193 based on the absence of host galaxy features
in the spectrum. In the only imaging attempt so far, \cite{2003A&A...400...95N}
did not detect the host galaxy in a 1200s R-band exposure.

PKS~0735+178 was classified and a BL Lac object by
\cite{1974ApJ...190L.101C}, who detected no emission lines, but a
strong $\lambda$2798\AA\ MgII absorption system at z=0.424 providing a
lower limit for the redshift. Later spectroscopy
\citep[e.g.][]{1993A&AS...98..393S,1994ApJS...93..125F,2001AJ....122..565R}
has confirmed the absorption system, but no emission features or host
galaxy absorption lines have been detected and the redshift remains
unknown. Optical imaging \citep{1988ApJS...66..361H,1991MNRAS.252..482A,
1993A&AS...98..393S,2000A&A...357...91F,2000ApJ...532..740S,2002A&A...381..810P}
has also failed to detect the host galaxy. \cite{2000ApJ...532..740S} give
a lower limit R $>$ 20.44 for the host galaxy based on a 440 s HST image through
the F702W filter.

Despite of the strong \citep[equivalent width $W_{obs}$ $\sim$
  3.6\AA,][]{2001AJ....122..565R} MgII absorption line, the absorbing
galaxy has remained hidden.  \cite{1988ApJS...66..361H} reported a
bridge of R-band emission from PKS~0735+178 towards the galaxy 7
arcsec NW with a SB of 26.0-25.2 mag/sq. arcsec. Later imaging by
\cite{1993A&AS...98..393S} did not see this bridge and the galaxy 7
arcsec NW was measured to have z = 0.645. There are two possible
detections of the absorbing galaxy: \cite{1999ASPC..159..385P}
reported a faint (R = 21.9) elongated structure 3 arcsec NE of
PKS~0735+178 and \cite{2000A&A...357...91F} saw faint emission 3.5
arcsec E of 0735+178 suggesting it could be the intervening z=0.424
system, although they couldn't rule out the possibility that it is
just a background target.

The brightness of the BL Lac nucleus relative to the host galaxy often
makes it very difficult to determine the host galaxy parameters
(magnitude and effective radius) accurately or to detect the host
galaxy in the first place. Given that the nucleus is often highly
variable, it is advantageous to wait for the moment when it is in a
low state. We are currently performing optical monitoring of $\sim$60
blazars, including RGB~0136+391 and PKS~0735+178
\citep[see \footnote{http://users.utu.fi/$\sim$kani/1m/index.html} and
  e.g.][for results of this monitoring]{2011A&A...530A...4A}.  We are
running a parallel program at the Nordic Optical Telescope (NOT) to
perform deep I-band imaging of a sample of gamma-ray emitting blazars
or potential gamma-ray emitters whenever our monitoring shows that the
optical brightness has fallen below a pre-defined threshold.  Instead
of observing the whole sample at once we wait for the optimal
moment. To improve the chances of detecting the host galaxy the
threshold level is chosen to represent a very low optical level of
each target considering the historical light curve.

In December 2008 the R-band brightness of RGB~0136+391 was observed to
be close to the lowest level since the beginning of our monitoring
(fall 2002) and in January 2011 the brightness of PKS~0735+178 reached
R = 16.7, a very low level considering the long-term historical light
curve \citep{2007A&A...467..465C}. Observations were promptly triggered
and performed at the NOT and the results are reported in this paper.

Throughout this paper we use the cosmology $H_0 = 70$ km s$^{-1}$
Mpc$^{-1}$, $\Omega_{M}$ = 0.3 and $\Omega_{\Lambda}$ = 0.7.

\section{Observations and data reduction}

\begin{figure}
\begin{center}
\epsfig{file=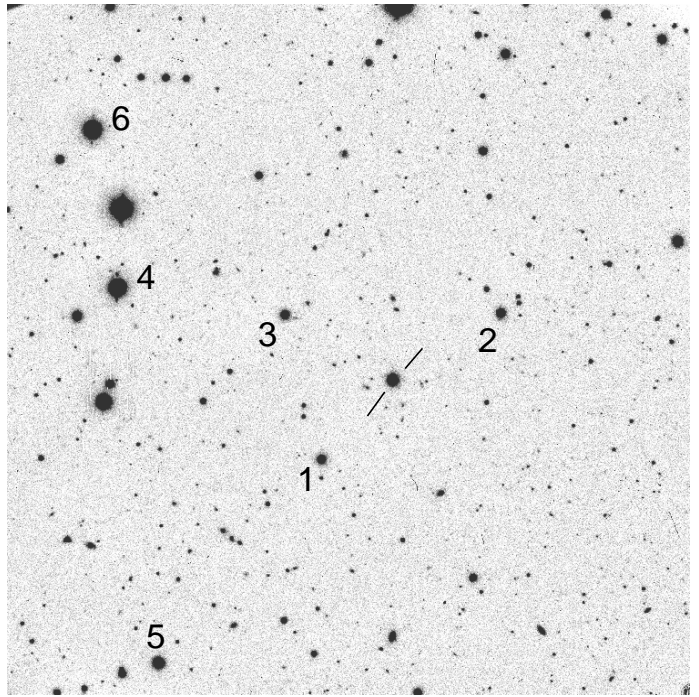,width=9cm}
\end{center}
\caption{\label{0136kentta} 
Summed image of RGB~0136+391 with the stars discussed in the text
labeled. Field size is $5.7 \times 5.8$ arcmin. North us up and east
is to the left.}
\end{figure}

\object{RGB 0136+391} was observed at the Nordic Optical Telescope
(NOT) through an I-band filter with almost uniform transmission
between 725 and 825 nm on December 7th, 2008. We used the ALFOSC
camera with a gain of 0.726 e$^-$/ADU and readout noise of 4.2
e$^-$. The field of view of the $2048^2$ E2V chip was $\sim 6.5 \times
6.5$ arcmin (0.189 arcsec/pixel).  Altogether 131 images with
individual exposure times between 25 - 100s were obtained resulting in
a total exposure time of 3830s. In order to make a fringe correction
image the telescope was moved between each exposure. The images were
bias-subtracted and flat-fielded with twilight flats and a fringe
correction image was constructed from the exposures and
subtracted. The images were then registered using 15 unsaturated stars
in the field and coadded.  The summed image shown in
Fig. \ref{0136kentta} has a FWHM of 0.61 arcsec.  Due to the presence
of a quarter Moon in the sky the background brightness was 19.2
mag/sq. arcsec, $\sim$ 0.8 mag brighter than average dark sky on La
Palma\footnote{La Palma technical note 115,
  http://www.ing.iac.es/Astronomy/observing/conditions/skybr/skybr.html}.
The field was calibrated using I-band observations made with the 35cm
KVA (Kungliga Vetenskapsakademien) telescope on La Palma in
photometric conditions on Jan 11th, 2012.  The standard star field
\object{PG 0231+055} \citep{1992AJ....104..340L} was observed at the
same airmass as the RGB 0136+391 field and the I-band magnitudes of
stars 1-6 were derived from these observations. The same stars were
then used to determine the zero point of our summed image. The error
of the derived zero point is 0.02 mag, excluding color effects (see
below).

\begin{figure}
\begin{center}
\epsfig{file=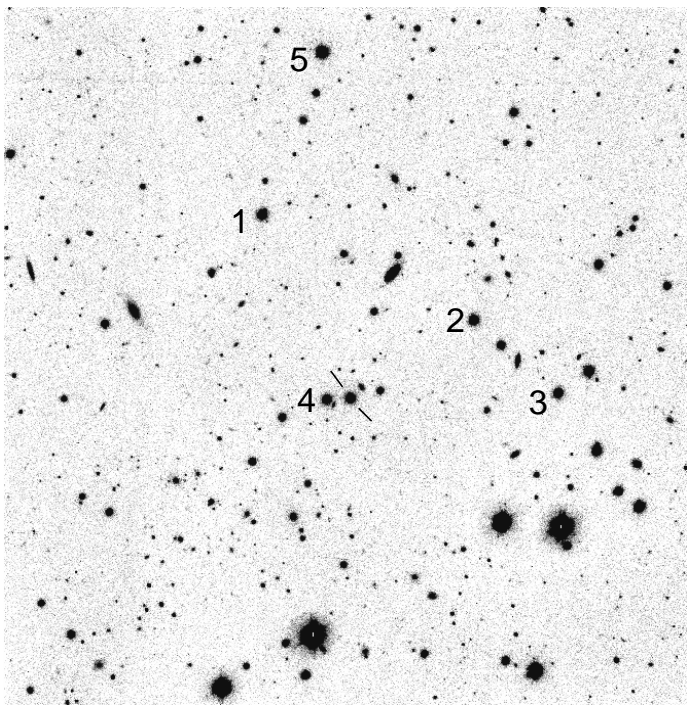,width=9cm}
\end{center}
\caption{\label{0735kentta} Summed image of PKS~0735+178 with the
  stars discussed in the text labeled. Field size is $5.2 \times 5.4$
  arcmin. North us up and east is to the left.}
\end{figure}

\object{PKS 0735+178} was observed with the same instrumental setup at
the NOT on January 18th, 2011. Altogether 55 images with exposure
times of 150 - 300 s were obtained. The images were processed in a way
identical to RGB~0136+391. The final coadded image shown in
Fig. \ref{0735kentta} has a total exposure time of 10250s and a FWHM
of 0.60 arcsec. The observations were influenced by almost full
($\sim$ 90\%) Moon 21 degrees away from PKS~0735+178. As a result the
sky background is relatively high (17.7 mag/sq. arcsec) and there is
scattered light present in the raw images resulting in an uneven
background.  The fringe removal procedure mostly eliminates the
scattered light but leaves a small (0.1\% peak-to-peak amplitude)
undulation of the background at large spatial scales, barely visible
in Fig. \ref{0735kentta}. Therefore, the background of PKS~0735+178
and its PSF stars was subtracted by fitting a two-dimensional linear
slope to the sky pixels surrounding the targets, effectively removing
any remaining tilt in the background.  Calibration of the field was
done through star 5, for which \cite{2007A&A...467..465C} give I =
$15.12 \pm 0.06$. This star was unsaturated in several exposures, which
were used to derive I-band magnitudes for stars 1-3, which were then
used to determine the zero point of the coadded image.

Since the transmission curve of the filter used here closely matches
the I-band transmission \citep{1990PASP..102.1181B}, the color effects
are expected to be small. According to the NOT zero point monitoring
pages\footnote{http://www.not.iac.es/instruments/alfosc/zpmon/} the
(V-I) color term is -0.03. Two of the stars used for calibration, star
4 in RGB~0136+391 and star 5 in PKS~0735+178, have V-I $\approx$ 0.8
\citep[][present work]{2007A&A...475..199N,2007A&A...467..465C}.
Assuming the rest of the stars used for calibration have similar
colors and that the host galaxy V-I = 2.6 \citep[elliptical galaxy
  color at z = 0.5,][]{1995PASP..107..945F} we derive the error of
host galaxy magnitudes due to filter mismatch to be $\approx$ 0.05 mag.

\section{Analysis}

We searched for evidence of the host galaxy by fitting two-dimensional
models to the observed image.  The detections are further tested by
simulations, which are also used to derive error bars for the derived
parameters.

The fitting procedure is the same as employed in our previous studies
\citep[e.g.][]{1999PASP..111.1223N,2008A&A...487L..29N}.
In short, the two-dimensional model consists of two components, the
unresolved core and a host galaxy whose surface brightness profile
follows the de Vaucouleurs law \citep[$\beta = 1/n = 0.25$, where $n$
  is the profile index from][]{1968adga.book.....S}.  The model is
described by five parameters: location $x$-$y$, assumed to be the same
for the core and the host galaxy, core magnitude $I_{\rm core}$, host
galaxy magnitude $I_{\rm host}$ and host galaxy effective (half light)
radius $r_{\rm eff}$. Since the host galaxies were known from previous
imaging to be very faint, the ellipticity and position angle were kept
at constant value = 0. We used a Levenberg-Marquardt loop to find the
set of parameters which minimized the $\chi^2$ between the data and
the model. The models were convolved with an empirical point spread
function (PSF) derived from stars in the vicinity of the target. In
the case of RGB~0136+391 we used stars 1-4 and in the case of
PKS~0735+178 stars 1-3.  The fitting proceeded in two
phases. First we fitted a model consisting of an unresolved component
only, i.e. the model had three free parameters ($x$, $y$ and $I_{\rm
  core}$). After this the position was fixed and the full core + host
galaxy model was fit with $I_{\rm core}$, $I_{\rm host}$ and $r_{\rm
  eff}$ as free parameters.

If the host galaxy is below detection limit these fits usually
converge towards $r_{\rm eff} \rightarrow 0$ or $r_{\rm eff}
\rightarrow \infty$.  In case we found $r_{\rm eff} > 0$ we tested the
significance of the detection by performing fits to simulated data.
We created a set of $\sim$100 simulated images of the target using the
fitted core and host galaxy values. The $\beta$ value for the surface
brightness profile of the simulated galaxy was drawn from a Gaussian
distribution with average equal to 0.25 and $\sigma = 0.025$. The
fitting, however, was made with fixed $\beta = 0.25$.  Each model was
convolved with a PSF consisting of a Moffat $\beta = 2.5$ profile with
a FWHM equal to the observed one, an ellipticity randomly drawn from
an uniform distribution between 0.0 and 0.06 and a position angle
drawn from an uniform distribution between 0.0 and 180 degrees.  To
these images we added Gaussian noise representing photon and readout
noise, scaled to the same level as in the observed data and a random
offset drawn from a Gaussian distribution with $\sigma = \sigma_{\rm
  sky} / 30$, where $\sigma_{\rm sky}$ is the sky rms scatter,
representing the error made in sky level estimation.

For each simulation we also constructed a second PSF, which slightly
deviated from the PSF used in convolving the model. The ellipticity of
the second PSF was drawn from an uniform distribution with mean equal
to the ellipticity of the first PSF and maximum deviation of
0.02. Similarly, the position angle deviated maximum 10 degrees from
the position angle of the first PSF.  To this second PSF we also added
random photon and readout noise plus an offset the simulate the sky
determination error. The second PSF was then used in the simulated
fits as the assumed PSF. Using a different PSF for the model creation
and for the fitting enabled us to roughly simulate the PSF variation
seen in the images and to produce residuals which qualitatively
resemble the residuals seen in actual images, although the latter tend
to be a bit more complex than what these simple simulations can
produce. The limits to the distributions above were selected to match
the residuals quantitatively, i.e. the peak-to-peak residuals in the
simulations were on average equal to the real residuals.  Similarly,
the average $\chi^2$ in the simulations was close to the $\chi^2$ of
the fit to the real data.

After the simulations we computed the standard deviation of the host
galaxy magnitudes $\sigma_{\rm host}$ and the range of effective radii
which includes 67\% of the simulated values. We didn't use standard
deviation for the effective radius because the distribution was
asymmetric with a longer tail towards high values. To consider a host
galaxy detected we required $\sigma_{\rm host} < 0.3$ mag and that the
median effective radius of the simulations is $> 2 \sigma_{\rm reff}$,
where $\sigma_{\rm reff}$ is the lower 67\% error bar.  The more
relaxed criterion for $r_{\rm eff}$ was used because $r_{\rm eff}$ is
generally more difficult to constrain and requiring $r_{\rm eff} > 3
\sigma_{\rm reff}$ would exclude host galaxies whose surface
brightness profile clearly exceeds any reasonable PSF variability.

\section{Results}

\subsection{RGB~0136+391}

\begin{figure}
\begin{center}
\epsfig{file=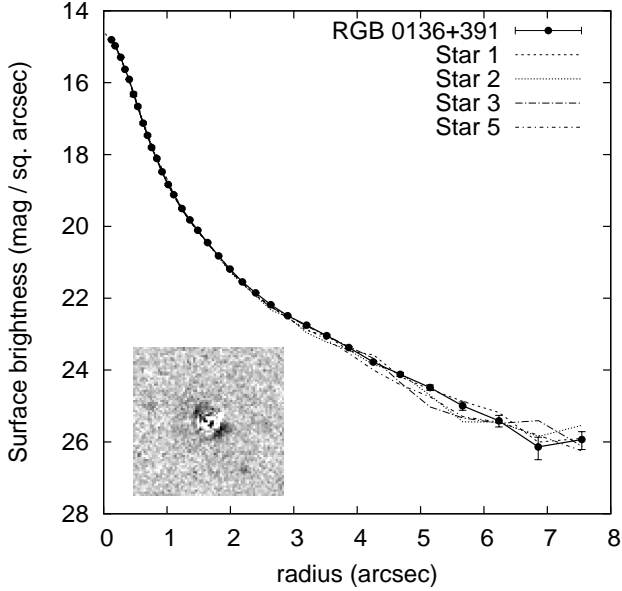,width=9cm}
\end{center}
\caption{\label{0136prof} I-band surface brightness profile of
  RGB~0136+391 (thick solid line) and four stars in the field. The
  stars are labeled in Fig. \ref{0136kentta}.  {\em Insert}: Residuals
  after subtracting the PSF from RGB~0136+391. Field size is
  14$\times$14 arcsec. The error bars include contribution from the
  empirical scatter in each bin and the error in sky background
  determination.  }
\end{figure}

Figure \ref{0136prof} shows the surface brightness (SB) profile of
RGB~0136+391 together with the profiles of four stars in the FOV. The
profiles of the stars have been scaled to the same total flux as
RGB~0136+391. The SB profile of RGB~0136+391 is totally consistent
with a point source, with no excess in the outer part of the profile
to indicate the presence of a host galaxy. Examination of the
two-dimensional residuals (Fig. \ref{0136prof}, insert) yields the same
conclusion: no obvious signs of the host galaxy above PSF residuals
can be seen.  The model fits confirm this conclusion: the final values
returned by the fits are $I_{\rm core} = 15.20 \pm 0.01$, $I_{\rm
  host} = 19.46 \pm 0.32$ and $r_{\rm eff} = 1.5^{+2.5}_{-0.8}$
arcsec, where the error bars come from the simulations described
above. This target thus remains unresolved.

We then derived a lower limit for the redshift of RGB~0136+391 using
the following procedure: if the distribution of BL Lac host galaxy
magnitudes is as shown by \cite{2005ApJ...635..173S}, then 95\% of BL
Lac host galaxies are brighter than $M_R = -22.0$. We thus created 100
simulated BL Lac objects at z = 0.25, 0.30, ..., 0.45 with core
magnitude $I_{\rm core} = 15.20$ and $I_{\rm host}$ and $r_{\rm eff}$
corresponding to $M_{\rm R}^{\rm host} = -22.0$ and $R_{\rm eff} = 10$
kpc. We first computed the absolute I-band magnitude $M_I$ using R - I
= 0.70 \citep{1995PASP..107..945F} and the apparent I-band magnitude
$m_I$ from
\begin{equation}
\label{apparentm}
m_I = M_I + DM + K_I + A_I - E(z)
\end{equation}
where $DM$ is the distance modulus, $K_I$ is the I-band K-correction from
\cite{1995PASP..107..945F}, $A_I = 0.15$ is the galactic extinction
\citep{1998ApJ...500..525S} and $E(z) = 0.84 * z$ is the evolution
correction, computed using the PEGASE code \citep{1997A&A...326..950F}
by assuming galaxy formation 11 Gyr ago and passive evolution
thereafter.

At each redshift we then fitted the 100 model images using the same
procedure as in the error simulations and computed the standard
deviation of $I_{\rm host}$ and $r_{\rm eff}$. The result of these
simulations was that at z = 0.45 ($m_I$ = 19.52) the host galaxy became
unresolved using our criteria. Thus at $z \leq 0.40$ we would have
detected the host galaxy with $>$95\% probability and thus assign a
conservative lower limit z $>$ 0.40 for RGB~0136+391.

\subsection{PKS~0735+178}

\begin{figure}
\begin{center}
\epsfig{file=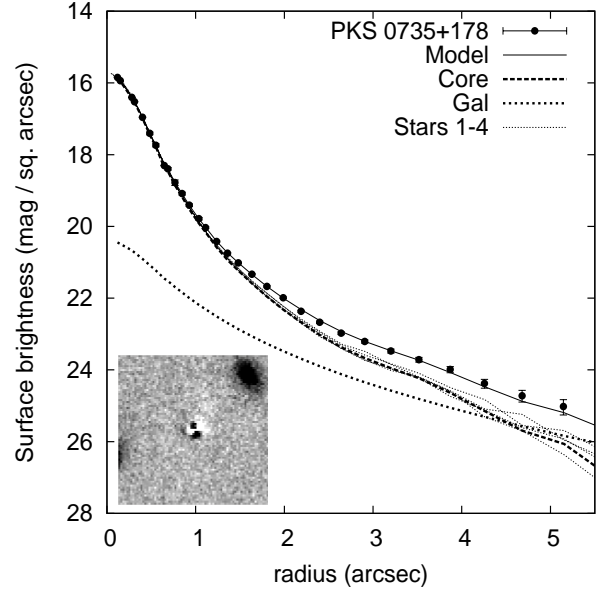,width=8.5cm}
\end{center}
\caption{\label{0735prof} Surface brightness decomposition of
  PKS~0735+178. In addition to PKS~0735+178 and model surface
  brightnesses, the profiles of stars 1-4 are shown. {\em Insert}: the
  residuals after subtracting the core + host galaxy model. Field size
  is 14.2 $\times$ 14.2 arcsec.  }
\end{figure}

Figure \ref{0735prof} shows the SB profile of PKS~0735+178 together
with stars 1-4. The stellar profiles have been again scaled to the
same total flux as PKS~0735+178. Contrary to RGB~0136+391, there is a
clear excess over the stellar profiles. Model fits yield $I_{\rm core}
= 16.19 \pm 0.01$, $I_{\rm host} = 18.64 \pm 0.07$ and $r_{\rm eff} =
1.8 \pm 0.4$ arcsec meeting our criteria for successful detection.
Adding the error from zero point determination and filter mismatch the
error of $I_{\rm host}$ becomes 0.11 mag.  Given the clear excess and
the fit results we conclude that we have detected the host galaxy or a
foreground galaxy (see Sect. \ref{juttua}).

The error bars from the simulations do not take into account possible
systematic errors, which are difficult to predict and/or quantify. We
have tested the sensitivity of our result with respect to our
particular choice of software by performing the fits with the {GALFIT}
software \citep{2002AJ....124..266P}. We used the same image, PSF and
sky background as before with the only difference being the
software. The result from this fit was $ I_{\rm host} = 18.57$
and $r_{\rm eff} = 2.0$ arcsec, i.e. the host galaxy magnitude
differs by 7\% and the effective radius by 11\% from our result. The
two programs thus provide very consistent results and we do not expect
the systematic errors due to software choice to dominate over
statistical noise.

Assuming that the detected galaxy is the host galaxy we estimated the
redshift of PKS~0735+178 using the results of \cite{2005ApJ...635..173S}.
Using Eq. \ref{apparentm} with $A_I = 0.068$ \citep{1998ApJ...500..525S}
and $M_I = -22.8$ we iteratively determined the the redshift consistent
with the observed I-band magnitude to be z = $0.45 \pm 0.06$, where
the error bar includes the contribution from host magnitude error and
the uncertainty of the method itself.

\section{\label{juttua}Discussion of PKS~0735+178}

\cite{2000ApJ...532..740S} derived an upper limit $m_R$ $>$ 20.44 for
the host galaxy, which together with our result implies R-I $>$ 1.7.
This is redder than the expected R-I = 1.2 for ellipticals at z=0.5
\citep{1995PASP..107..945F}, possibly an indication of systematic
differences between the two studies, or of dust reddening in the host
galaxy. If significant dust absorption is present in PKS~0735+178,
our redshift estimate is too high.

\begin{figure}
\begin{center}
\epsfig{file=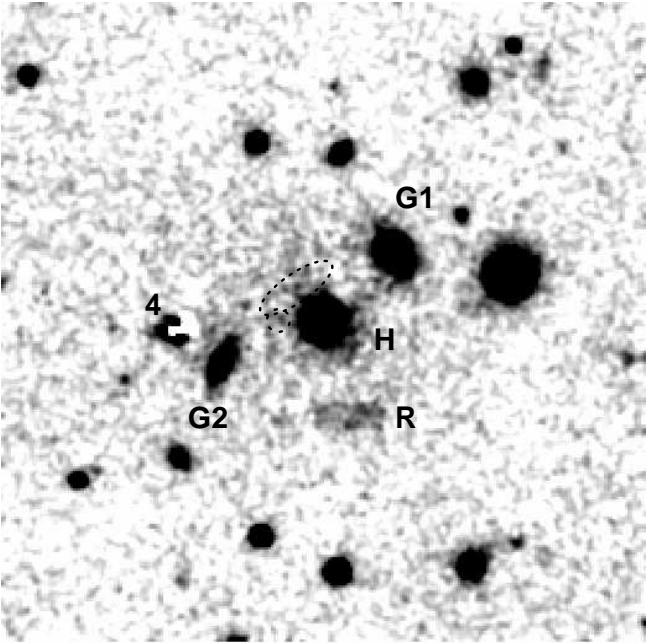,width=8.5cm}
\end{center}
\caption{\label{0735zoom} 47.4 $\times$ 47.4 arcsec field around
  PKS~0735+178 after subtracting the scaled PSF from
  PKS~0735+178 and star 4.  The image has been smoothed with a
  Gaussian kernel with $\sigma$ = 1.0 pix.  Galaxies G1 and G2 and the
  rectangular test feature R are discussed in the text. H is the host
  galaxy of PKS~0735+178. The dotted ellipse and circle mark the
  location of the features discussed by \cite{1999ASPC..159..385P} and
  \cite{2000A&A...357...91F}, respectively. }
\end{figure}

Using the I-band data obtained here and previous R-band data we now
discuss the identity of the MgII absorber. Galaxy G1 in
Fig. \ref{0735zoom} can be directly ruled out due to the redshift z =
0.645 measured by
\cite{1993A&AS...98..393S}. \cite{1999ASPC..159..385P} and
\cite{2000A&A...357...91F} detected weak R-band emission 3\farcs0 NE
and 3\farcs5 E of PKS~0735+178, respectively, and discussed the
possibility of these features being the MgII absorber. Neither of
these emission features are visible in our PSF-subtracted I-band image
(Fig. \ref{0735zoom}).  Since our images were obtained in bright
moonlight and thus not optimal for detecting faint diffuse emission,
we tested our sensitivity with the following procedure.

Firstly, since \cite{2000A&A...357...91F} do not give a magnitude for
the feature they detected, we cannot estimate if it was detectable in
our image. We therefore concentrate on the elongated feature reported
3\farcs0 NE of PKS~0735+178 by \cite{1999ASPC..159..385P}.  To roughly
simulate this feature we created a rectangular test object marked with
R in Fig. \ref{0735zoom}. Note that we did not create this feature at
the correct location to better compare with the actual image. The
feature has a uniform I-band surface brightness of 24.3
mag/sq. arcsec and its dimensions and integrated magnitude 21.9
correspond to those of the elongated featured reported by
\cite{1999ASPC..159..385P}. The test feature is just visible over the
background noise and thus we should have detected the
\cite{1999ASPC..159..385P} feature in our image if its R-I color $>$
0. Since the range of R-I colors of E to Im galaxies at z = 0.424 is
expected to be form 1.1 to 0.4, respectively
\citep{1995PASP..107..945F}, we conclude that we should have detected
the \cite{1999ASPC..159..385P} feature, if its colors are within the
range of ``normal'' galaxies. Given the non-detection in our I-band
image the \cite{1999ASPC..159..385P} feature is unlike to be the
absorber. It could be a line emitting cloud whose emission wavelength
falls into the R-band but outside the I-band. Of the lines bluewards
of the R-band, $\lambda$5007 [OIII] is the most likely candidate. This
line would place the cloud at z = 0.13-0.44 when 50\% transmission
limits of the R-band filter are considered.

Since we do no detect the feature discussed by
\cite{1999ASPC..159..385P} and the detection probability of the
\cite{2000A&A...357...91F} feature cannot be estimated we turn our
attention to galaxy G2. This galaxy lies at projected distance of
8\farcs2 (46 kpc) from PKS~0735+178. Given the rest frame equivalent
width of the MgII line in PKS~0735+178, $W_0$ = 3.6\AA$/(1+z_{abs}) =
2.5$\AA, we ask if this galaxy could be the absorber. For comparison
we use \cite{2007ApJ...658..161Z}, who studied 2844 MgII absorbers
with $W_0 > 0.8$\AA\ and $0.37 < z < 1$ using the SDSS DR4. We first
note that $W_0 = 2.5$\AA\ implies very strong absorption: 93\% of the
systems studied by \cite{2007ApJ...658..161Z} have $W_0 < 2.5$\AA. If
we concentrate on the relevant section of the parameter space in
\cite{2007ApJ...658..161Z}, $0.37 < z < 0.55$ and $W_0 > 1.58$\AA, we
note that MgII absorbers can be found up to projected distances of
$\sim$ 100 kpc from the background source. In this section of the
parameter space the probability of finding an MgII absorber at
projected distance of 46 kpc is $\sim$ 25\% \citep[see Fig. 9
  in][]{2007ApJ...658..161Z}.  Given that the probability of finding
an Mg II absorber at a given distance decreases with $W_0$ and the
$W_0$ of the MgII line in PKS~0735+178 is at the high end of the
distribution, the probability of G2 being the absorber is probably
considerably lower.  The I-band magnitude of G2 is $20.4 \pm 0.1$
which gives R-I = 0.5 when combined with the R-band magnitude in
\cite{1999ASPC..159..385P}.  This color is consistent with a late type
(Scd) galaxy at z = 0.424.  The absolute magnitude of G2 at z = 0.424
would be $M_R = -20.9 \approx M^*_R + 0.3$.  In the light of these
results G2 cannot be completely ruled out as the absorbing galaxy,
although finding such a system is relatively unlikely.

As the next option we consider the possibility that the detected host
galaxy is actually a foreground object and not the real host galaxy.
If this is the case, then the impact parameter must be very close to
zero since the detected galaxy appears to be well centered on the BL
Lac nucleus.  To check for possible offsets between the core and the
host galaxy we performed the host galaxy fit with 7 free parameters,
i.e. in addition to $I_{\rm core}$, $I_{\rm host}$ and $r_{\rm eff}$
also the x-y positions of the core and the host galaxy were allowed to
change freely. The core - host galaxy offset was 0.67 pixels (0.13
arcsec) in this case, indicating that the core is well centered on the
host galaxy. Such a system is capable of producing gravitational
lensing phenomena, but \cite{2003AJ....125.2447R} did not detect any
signs of gravitational macrolensing (Einstein rings or multiple
images) in their 8.46 and 14.94 GHz maps with $\sim$ 0.2 arcsec
resolution. Furthermore, the result of \cite{2007ApJ...658..161Z}
indicate that strong absorbers are associated with galaxies with
intense star forming, which should be visible in the optical spectrum
as narrow emission lines. These findings do not disprove the
hypothesis that the detected galaxy is an intervening system with a
small impact parameter, but they make it less attractive with respect
to the alternative, that we have detected the actual host galaxy.

Lastly we note that the derived redshift $0.45 \pm 0.06$ is within
errors equal to the absorption redshift 0.424. It could thus be
possible that the absorption occurs in the host galaxy of PKS~0735+178
itself. This would imply the PKS~0735+178 host galaxy to be a gas-rich
system since strong Mg II absorption is often associated with damped
Lyman-alpha systems (DLAs) with high HI gas content
\citep{2004MNRAS.352.1291P}.  Further observations are clearly needed
in order to determine which, if any, of the above scenarios is the
correct one. A spectrum of G2 is needed in order to see if its
redshift is compatible with being the absorbing galaxy.  An
ultraviolet spectrum of PKS~0735+178 detecting the Lyman-alpha line
would also shed new light on the nature of the absorber by measuring
of the HI column density.
 
\section{Conclusions}

We have presented deep I-band imaging of two BL Lacertae objects,
RGB~0136+391 and PKS~0735+178, obtained during a deep minimum in the
optical light curve. Despite of a deep exposure obtained under
excellent seeing conditions (0\farcs6) the host galaxy of RGB~0136+391
remains unresolved. We derive z $>$ 0.40 for RGB~0136+391 using
simulated host galaxies at different redshifts with $M_R = -22.8$ and
$R_{\rm eff} = 10$ kpc. This makes RGB~0136+391 one of the most
distant VHE $\gamma$-ray sources detected up to date.

The surface brightness profile of PKS~0735+178 reveals an excess
over a point source, which, if fitted by a de Vaucouleurs profile,
corresponds to a galaxy with I = $18.64 \pm 0.11$ and $r_{\rm eff} =
1.8 \pm 0.4$ arcsec. Under the assumption that BL Lac host galaxies
can be used as standard candles \citep{2005ApJ...635..173S}
we derive z = $0.45 \pm 0.06$ for PKS~0735+178.

We also discuss the identity of the galaxy responsible for the strong
MgII absorption line in PKS~0735+178 at z$_{abs}$ = 0.424 using
previous R-band imaging and the I-band data in this paper. We do not
detect the two faint objects suggested earlier to be responsible for
the absorption or any other obvious candidates in the vicinity of
PKS~0735+178. We identify a galaxy 8\farcs2 of PKS~0735+178 (46 kpc
projected distance) as a candidate for the absorber, although the
probability of strong absorption ($W_0$ = 2.5\AA) at this projected
distance is very low. Further observations are needed in order to
reveal the identity of the absorber towards PKS~0735+178.

\acknowledgements 

The data presented here were obtained with ALFOSC, which is provided
by the Instituto de Astrofisica de Andalucia (IAA) under a joint
agreement with the University of Copenhagen and NOTSA.  This research
has made use of NASA's Astrophysics Data System Bibliographic
Services.  This research has made use of NASA's Astrophysics Data
System Bibliographic Services. This research has made use of the
NASA/IPAC Extragalactic Database (NED) which is operated by the Jet
Propulsion Laboratory, California Institute of Technology, under
contract with the National Aeronautics and Space Administration.

\bibliographystyle{aa}

\bibliography{19848.bib}

\end{document}